# Solar Eruptive Events (SEE) 2020 Mission Concept

R. P. Lin, A. Caspi, S. Krucker, H. Hudson, G. Hurford, (SSL/UCB)
S. Bandler, S. Christe, J. Davila, B. Dennis, G. Holman, R. Milligan, A. Y. Shih (GSFC)
S. Kahler (AFRL); E. Kontar (Glasgow); M. Wiedenbeck (JPL); J. Cirtain (MSFC)
G. Doschek, G. H. Share, A. Vourlidas (NRL); J. Raymond (SAO); D. M. Smith (UCSC)
M. McConnell (UNH); G. Emslie (WKU)

**Abstract** Major solar eruptive events (SEEs), consisting of both a large flare and a near simultaneous large fast coronal mass ejection (CME), are the most powerful explosions and also the most powerful and energetic particle accelerators in the solar system, producing solar energetic particles (SEPs) up to tens of GeV for ions and hundreds of MeV for electrons. The intense fluxes of escaping SEPs are a major hazard for humans in space and for spacecraft. Furthermore, the solar plasma ejected at high speed in the fast CME completely restructures the interplanetary medium (IPM) - major SEEs therefore produce the most extreme space weather in geospace, the interplanetary medium, and at other planets. Thus, understanding the flare/CME energy release process(es) and the related particle acceleration processes are major goals in Heliophysics. To make the next major breakthroughs, we propose a new mission concept, SEE 2020, a single spacecraft with a complement of advanced new instruments that focus directly on the coronal energy release and particle acceleration sites, and provide the detailed diagnostics of the magnetic fields, plasmas, mass motions, and energetic particles required to understand the fundamental physical processes involved.

## 1 Introduction

Solar eruptive events (SEEs) consist of both a large flare and a near simultaneous large, fast, coronal mass ejection (CME). The largest SEEs are the most powerful explosions in the solar system, releasing $10^{32}$-$10^{33}$ ergs on time scales as short as tens of minutes. They are also the most powerful and energetic particle accelerators in the solar system, producing ions up to tens of GeV and electrons up to hundreds of MeV. For flares, the accelerated particles often contain 10-50% of the total energy released in the flare, a remarkable efficiency that indicates the particle acceleration and energy release processes are intimately related. Similar processes appear to occur elsewhere in the universe, in stellar flares, magnetars, etc. Escaping solar energetic particles (SEPs) appear to be accelerated by the shock wave driven by the fast CME at altitudes of ~2-40 $R_s$, with an efficiency of ~10%, about what is required for supernova shock waves to produce galactic cosmic rays. Thus, SEEs are our most accessible laboratory for understanding the fundamental physics of transient energy release and particle acceleration in cosmic magnetized plasmas. The intense fluxes of SEPs are a major hazard for humans in space and for spacecraft. Furthermore, the ~$10^{15}$–$10^{16}$ g of solar plasma ejected at high speed in a fast CME significantly restructures the interplanetary medium (IPM) - major SEEs therefore produce the most extreme space weather in geospace, the IPM, and at other planets. Thus, understanding the fundamental energy release process(es) in SEEs and the related particle acceleration processes are major goals in Heliophysics.

Our basic working model starts with the buildup of energy in the magnetic field of the solar corona over days to weeks. The free magnetic energy is released explosively through a process that is only poorly understood, but most probably involves magnetic reconnection to enable the magnetic configuration to relax to a lower energy state. The magnetic energy so released appears primarily in the form of the acceleration of particles (both electrons and ions), the acceleration (mass motion) and heating of the bulk material of the CME, and heating of the flare plasmas. Subsequently, further particle acceleration occurs at the CME-driven shock. The detailed partition of the released energy among these various products is shown in Fig. 1 for the SEE on 2002 July 23 (Emslie et al. 2004, 2005). In this case, and in a few other well-observed events, there is an approximate equipartition of energy between the flare and the CME.

Great progress has been made in developing this picture of SEEs, but fundamental science questions remain unanswered.

- How is magnetic energy suddenly released to produce both a flare and a CME?
- How are CMEs accelerated?
- How can flares accelerate electrons and ions so efficiently?
- How are escaping SEPs accelerated?
- How are the flare and the CME plasmas heated?
- How is magnetic energy built up and stored in the corona prior to an SEE?

The answers to these and other questions are needed before a comprehensive physical understanding can be claimed or reliable forecasting made of these events. This will require a coordinated program of theory and modeling of the fundamental physical processes combined with our proposed SEE 2020 space mission and supporting space- and ground-based observations.

Given our current state of knowledge and the development of new instrumentation capable of advanced observations of the most intense SEEs, we believe that the next solar maximum in the 2020 time frame presents an excellent opportunity to make dramatic advances in this field. Since the largest events are of great scientific interest and the most important for space weather, they must be the focus of the effort. Consequently, all instruments must be optimized for observations of the largest M- and X-class events while maintaining high sensitivity, and they must have coordinated and complementary capabilities to maximize the science return from simultaneous observations. The instruments must have the resolutions and dynamic ranges in space, time, and energy/wavelength required to resolve the weakest features of interest in the presence of the strongest components, without saturation. The observations of the largest events will thus provide the greatest information possible while at the same time allowing sensitive and accurate observations of many weaker events.

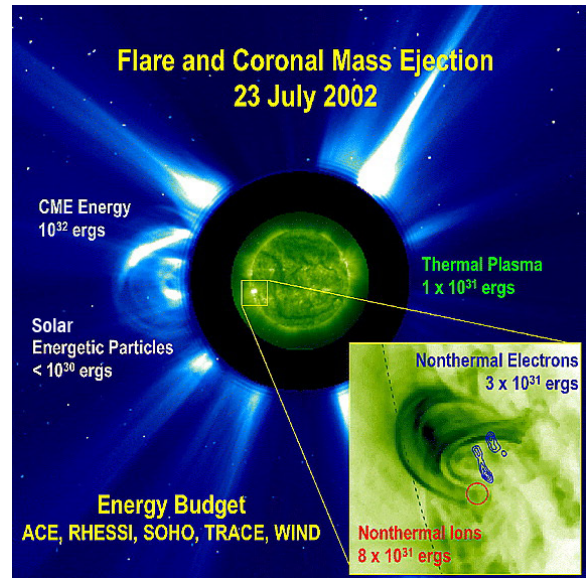

Fig. 1. Images and energy budget of the major components of an SEE (Emslie et al. 2004). This comprehensive picture required input from instruments on 5 individual spacecraft.

The basic philosophy of the proposed SEE 2020 research program is to concentrate on the sites of the energy release and particle acceleration in the solar corona, and to characterize the most energetically important components as they evolve in time and space. Observations will be made of these sites before, during, and subsequent to the event to characterize the magnetic fields, the plasma density, temperature, and flow velocities, and plasma wave fields (where possible), as well as the accelerated particles. In addition, the accelerated particles, mass motions, and heated plasma will be observed as close to the release site as possible both in space and in time so that the important processes involved can be accurately inferred with minimal uncertainties due to propagation effects and temporal evolution.

### 1.1 Location and Nature of the Energy Release Site

Many lines of reasoning point to the corona as the location of the energy release site, and the coronal magnetic field as the source of the energy that powers these SEEs. Relatively weak coronal sources are sometimes detected. When two coronal X-ray sources with opposite energy gradients are observed one above the other, the energy release site can be inferred to lie between them (Fig. 2). In the first such observation, Sui and Holman (2003) found that the energy release site was at an altitude of between ~9 and 23 Mm. With RHESSI's limited dynamic range, weak coronal HXR sources are difficult to detect in the presence of the much brighter footpoint sources, but systematic studies of flares whose footpoints are occulted show that coronal HXR emission is commonly present. Yohkoh observations (e.g., the Masuda flare) and more recent RHESSI observations of an occulted flare (Krucker et al. 2010) suggest that the particle acceleration/energy release region is located about 5-10 Mm above the flare loops.

The acceleration vs. time profiles for fast CMEs up to ~4 $R_s$ are found to be closely synchronized with energy release of the associated large flares, as measured by the hard X-ray flux vs. time profile (Qui et al. 2004; Temmer et al. 2008, 2010). This is consistent with magnetic reconnection occurring in the current sheet behind the CME, and confirms that the energy release processes for CMEs and flares are intimately related.

## 2 SEE 2020 Required Measurements

### 2.1 Particle Acceleration

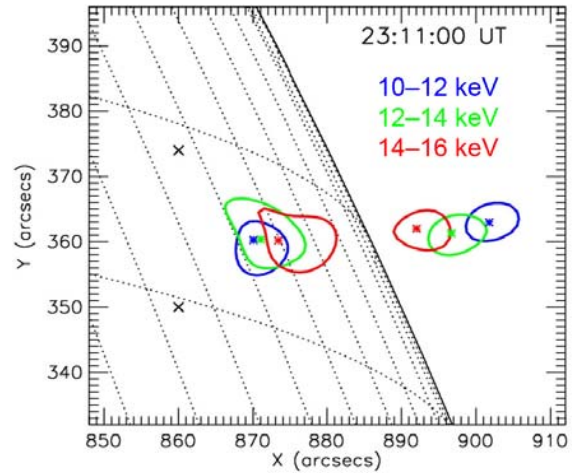

Fig. 2. RHESSI observation of two coronal thermal X-ray sources that are believed to bracket the flare energy release site. Each HXR footpoint is indicated with an ×.

The large fraction of the released magnetic energy that goes into the acceleration of particles, both electrons and ions – perhaps as high as 50% – makes the determination of the physical processes by which this is achieved central to any attempt at understanding SEEs. Accelerated particles that interact at the Sun are most effectively explored by observing the X-rays and gamma-rays that they produce. Detection of outwardly directed particles has been limited to *in situ* observations, but new techniques may now be possible to image those particles using energetic neutral atoms (ENAs) resulting from charge exchange. Such images could provide spatial and temporal connections to the shock regions imaged by coronagraphs.

#### 2.1.1 Flare-accelerated Electrons

The physical mechanisms that accelerate electrons and that affect their distribution as they propagate away from the acceleration site are still unknown. In large flares, the required rate of acceleration of electrons above 20 keV can exceed $10^{37}$ electrons s$^{-1}$. Such a large flux of charged particles imposes fundamental constraints on the global electrodynamics of the system and highlights the importance of collective effects. In the standard flare model, the nonthermal electrons are accelerated in the corona and stream down the field lines to the footpoints, where they interact collisionally leading to bright HXR emission. RHESSI observes electrons primarily where they are stopped in the thick target at the footpoints. Consequently, it is difficult to isolate properties of the acceleration mechanism because of unobserved transport effects such as collective plasma effects.

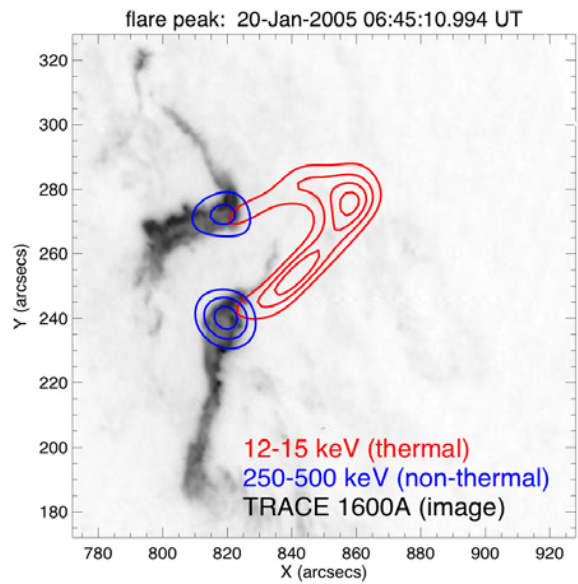

Fig. 3. Note the lack of HXR emission (blue) detected from the flare loop (red) at the peak time of the flare (Krucker et al. 2008).

As illustrated in Fig. 3, RHESSI's dynamic range limitations make it difficult to image X-rays from electrons in the tenuous corona in the presence of much brighter footpoints. Also, the low-surface-brightness albedo of HXRs back-scattered from the photosphere - crucial for electron anisotropy measurements - cannot be imaged with RHESSI. Both the instrument sensitivity and the dynamic range must be increased by a factor of at least 10, and preferably 100, to enable such observations to be made routinely. These improvements can be achieved with HXR grazing-incidence focusing optics that operate up to ~100 keV. HXR observations of electrons in the corona with these improved capabilities will show how the electron spectrum evolves in space and time from the acceleration site to the footpoints. The high sensitivity and dynamic range will also allow the HXR albedo to be imaged.

The degree of electron anisotropy can also be determined from HXR and gamma-ray imaging polarimetry. New polarimeters, designed for the first time specifically for this purpose, can provide measurements of the magnitude and direction of the polarization that can distinguish, for example, between different particle acceleration models.

### 2.1.2 Flare-accelerated Ions

Because of their large energy content, no model of SEEs can be complete without addressing the acceleration of ions. In fact, there appears to be a rough equipartition in energy between electrons and ions that interact with the Sun (Ramaty & Mandzhavidze 2000), and the flare acceleration of ions and electrons is generally tightly correlated (e.g., Shih et al. 2009). However, the sources of ion-associated emission and electron-associated emission in the two flares for which both were well observed have shown surprising spatial separations, indicative of unexplained differences in the acceleration and/or transport processes (Hurford et al. 2003, 2006). Fig. 4 shows the best example of this separation. The relationship between the locations where ions and electrons interact can be studied in a systematic fashion using observations with the sensitivity and resolution to detect and resolve sources in medium to large flares. This requires high-resolution gamma-ray imaging with a factor of >~10 increase in sensitivity, and <10 arcsec FWHM resolution (with a clean point-spread function) compared to RHESSI's 35 arcsec resolution. Imaging is necessary not only in the neutron-capture line as achieved with RHESSI, but also in the nuclear de-excitation gamma-ray lines (both resolved and unresolved) and in the positron-annihilation line at 511 keV. Comparing the spectral, spatial, and temporal behavior of ions and electrons requires X-ray and gamma-ray imaging spectroscopy from <10 keV to >15 MeV. Moderate-resolution gamma-ray spectroscopy (~3% FWHM) can disentangle the various gamma-ray components that are produced by energetic ions. Besides information about the spectrum, directivity, and composition of the accelerated ions, these observations provide information about the ambient plasma in the interaction region, including abundances.

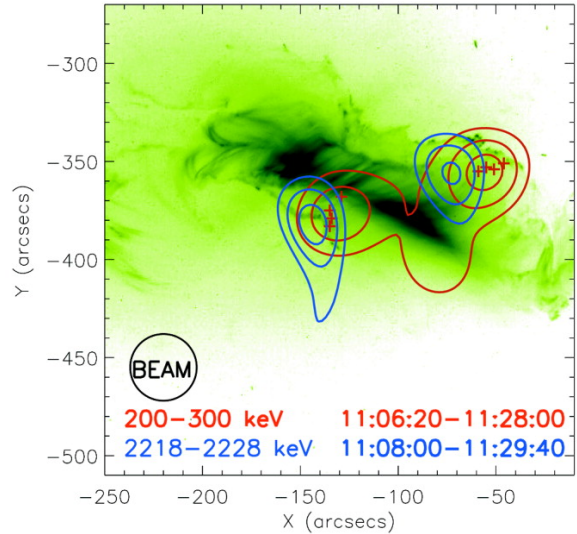

Fig. 4. Unexplained spatial separation of X-ray (red) and gamma-ray (blue) sources during 2003 Oct. 28 flare (Hurford et al. 2006).

Determinations of the total ion energy content have large uncertainties largely because the spectrum is completely unknown below a few MeV, where a major fraction of the total energy may reside. Three promising signatures of such low energy ions are, (1) radiative-capture gamma-ray lines (MacKinnon 1989; Share et al. 2001), (2) Doppler-shifted Lyman-alpha 1216 Å line emission produced by ~300 keV protons that charge exchange in flight, and (3) Doppler-shifted He II 304 Å emission produced by accelerated alpha particles.

### 2.1.3 Particle acceleration in CME-driven shocks

*In situ* observations of interplanetary SEPs will be made by inner-heliospheric missions such as Solar Orbiter, Solar Probe Plus, etc.. However, the broader understanding of the spatial and temporal origins of SEPs has been severely restricted by the limitation to such *in situ* measurements since SEPs are scattered in turbulent magnetic fields. *In situ* observations at only two or three separate locations from an evolving 3-D heliospheric shock cannot address such basic questions as the relative importance of perpendicular and parallel shocks (Tylka and Lee 2006), the radial dependence of acceleration efficiencies, and why some shocks fail to accelerate detectable SEPs (Cohen et al. 2005). These questions can only be addressed by remote imaging of SEP signatures produced by the interactions with ambient coronal and interplanetary particles (Kahler and Ragot 2009).

The serendipitous discovery of energetic neutral atoms (ENAs) at MeV energies associated with the 2006 December 4 eruptive event (Mewaldt et al. 2009, 2010) has introduced a new probe of SEP acceleration. ENAs are not deflected by the magnetic field, and the charge exchange that neutralizes the ions does not significantly affect their energies. It is unlikely that ENAs produced in the chromosphere can escape without charge exchanging again, but imaging of CME-associated ENAs in association with coronagraph CME observations can then address basic questions of shock characteristics and SEP acceleration. ENA measurements from ~10 keV to ~10 MeV include both higher-energy ENAs produced by charge exchange with heavy ions (e.g., oxygen) and lower-energy ENAs that are predicted to result from charge exchange with neutral hydrogen and hydrogen-like helium in the ambient medium.

High-sensitivity gamma-ray imaging will distinguish between interacting flare-accelerated ions that are localized to the flare loops and those ions accelerated by the CME shock that precipitate to form a larger, diffuse source. At particle energies above ~100 MeV, imaging of pion-decay gamma-ray emission may distinguish between flare-accelerated particles in the chromosphere and CME-shock-accelerated ions in the high corona or solar wind.

In addition to measuring the SEPs themselves, understanding the acceleration process requires measuring the shock characteristics, the plasma conditions, and the seed population from which they are accelerated. The coronagraph images described earlier track the latitudinal and radial development of the CME-driven shock at the crucial heights within ~4 $R_{sun}$, where the SEPs are believed to be accelerated. Spectroscopy with a UV coronagraph measures parameters of both the pre-shock and the post-shock coronal plasma, including the density, temperature, compression ratio, outflow speed, and the elemental and charge-state composition. Such observations with high sensitivity, cadence, and spectral range will also measure the seed population of suprathermal protons. The seed population may also consist of the flare-accelerated particles that would be simultaneously studied.

## 2.2 Evolution of SEE Magnetic Fields

Any attempt to understand SEEs must begin with knowledge of the coronal magnetic field, the configuration leading up to the energy release, and the evolution during and following the event. Since it is difficult to measure the coronal magnetic field ***strength***, estimates are traditionally made from force-free extrapolations of the measured photospheric vector magnetic field. Direct measurements of coronal magnetic fields with the Advanced Technology Solar Telescope (ATST) and from the Frequency Agile Solar Radiotelescope (FASR) should be available by 2020. These measurements can be coupled with measurements of the coronal magnetic field ***structure*** from space through high-spatial-resolution optical and EUV coronagraph observations. Images of the corona taken from the ground during recent eclipses (Fig. 5) demonstrate that fine-scale coronal structure in electron density is visible, and that it traces the magnetic field above the solar limb. An eclipse-like visible-light coronagraph would image the magnetic field without temperature bias, and can also observe the pre-eruption magnetic configuration as regularly done during eclipse observations. Such images taken in space every 10 s with ~5 arcsec angular resolution between 1.02 (14 Mm) to 4 $R_{Sun}$ would reveal the changes to the magnetic configuration on the expected fine spatial and temporal scales This range covers the expected altitude of the initial energy release. Extending coronagraph observations out to 15 $R_{Sun}$ will allow the energy release site to be observed in off-limb events and to follow the CME during its acceleration phase up to its full velocity. Such observations can provide strong constraints on theories of CME initiation and acceleration (see Forbes et al. 2006). These measurements can also determine the coronal magnetic field configuration during the build-up of energy before the event.

## 2.3 Evolution of SEE Plasmas

Measurements of plasma parameters in the SEE energy release/particle acceleration sites with appropriate spatial and temporal resolution are needed. It is necessary to obtain diagnostics of the current sheets where magnetic reconnection occurs. Spectra from an off-limb EUV imaging spectrometer can determine the density, temperature, inflow speeds (Yokoyama et al. 2001), outflow speeds (Wang et al. 2005), and turbulent velocities associated with large-scale current sheets that have been observed extending from the flare loops to the CME core.

### 2.3.1 Flare plasma heating

Observations of the co-evolution of energetic particles and thermal plasma are critical to obtaining an understanding of particle acceleration and plasma heating in solar flares. Direct plasma heating is expected to occur in the energy release region in the corona. The hottest flare plasma of one event has been located high in the corona, possibly at the location of particle acceleration (Caspi and Lin 2010), and could be evidence of such direct heating. It is well established, however, that accelerated electrons heat plasma away from the acceleration region, especially in the chromosphere and lower transition region. The chromospheric plasma indirectly heated in this way expands upward into the corona (Abbett & Hawley 1999; Allred et al. 2005), and this, in turn, changes the conditions in the acceleration region. Coincident imaging spectroscopy is required in the EUV, soft X-rays, and hard X-rays to obtain a clear picture of this co-evolution of the particle acceleration and plasma heating. The observations must cover a full active region with <10 s cadence and moderate spatial resolution so the same features are observed in the EUV, and soft and hard X-rays. High spectral resolution is desired to take advantage of computed line shapes, to better determine plasma velocities down to ~10 km s$^{-1}$, and determine the presence of line excitation by nonthermal particles. A high sensitivity EUV instrument is also critical for observing pre-impulsive phase energy release and plasma evolution.

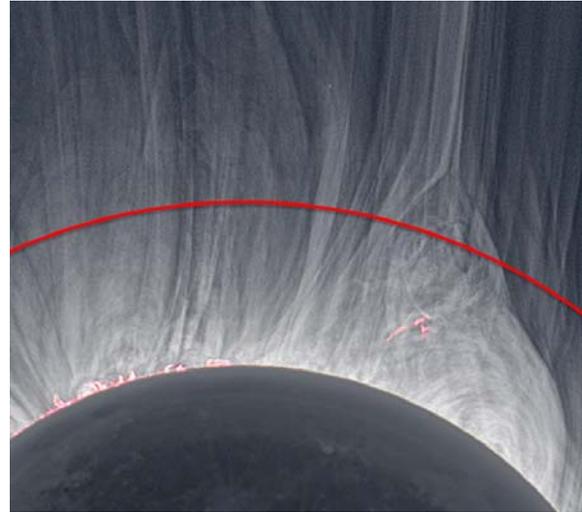

Fig. 5. Eclipse picture from Druckmüller et al. (2006) with ~5 arcsec effective resolution showing the complex magnetic structure not visible with STEREO/COR1 below 1.4 $R_{Sun}$.

### 2.3.2 CME plasma heating

Although the kinetic energy from the large-scale mass motion of the CME is the obvious manifestation of energy escaping from the Sun, the ejected plasma is strongly heated, as is apparent when erupting filaments change from absorption to emission features. Understanding the energetics of this aspect is crucial to understanding the total energy budget. The energy in the heated plasma is comparable to the kinetic energy of the CME in all the events studied so far (Landi et al. 2010). While heating from dissipation of the magnetic field is strongly suspected, the nature of the heating mechanism is not yet known. Detailed studies over many events of such implications can be made using UV and EUV spectroscopy of lines formed in the $10^4$ to $2\times10^6$ K range at several heights between ~1.2 and ~3 $R_{sun}$. To measure the heating as a function of height, one requires the spectral range and sensitivity to obtain good density diagnostics, along with observations at multiple heights and high enough cadence to identify specific features at multiple heights. One also needs a spectral range that provides a large number of lines because the plasma can quickly depart from ionization equilibrium and more constraints are needed to determine plasma parameters.

## 3 Mission Concept

The proposed SEE 2020 mission will measure all the energetically and diagnostically important aspects of solar eruptive events. It will have the temporal, spatial, and spectral resolution and coverage necessary to identify the coronal conditions needed for an eruption to take place and the physical processes that make it happen. The instruments necessary to make the required new measurements are listed in Table 1 and described in more detail in separate white papers. They will be combined together on a single spacecraft, likely in low-Earth orbit. Observations from ground-based observatories (such as ATST and FASR) and other space missions (such as Solar Orbiter, Solar Probe Plus, and Solar-C) will complement this mission concept by providing context information on magnetic fields, thermal plasma, and interplanetary SEPs. The technologies capable of achieving the instrument requirements that have not already been proven are currently under development and will be demonstrated on a variety of rocket and balloon missions over the next few years. We envision SEE 2020 to be similar in cost and schedule to the Solar Dynamics Observatory (SDO).

**Table 1.** Complement of instruments for the SEE 2020 mission to be flown on a single spacecraft, likely in low-Earth orbit.

| Instrument | Key Instrument Characteristics | New Measurement | Primary Science Question |
|---|---|---|---|
| White-light coronagraph | Angular resolution: 5 arcsec<br>Full Sun FOV: 1.02 to 12 $R_{sun}$ | Observe the rapidly changing structure of the magnetic field at the energy release site. | How does evolution of the coronal magnetic field lead to eruptive events? |
| Off-limb EUV imaging spectrometer | Angular resolution: 1 arcsec<br>Temperature: $10^4$–$10^7$ K<br>Cadence: ~10 s | Observe CME acceleration and heating at very low altitudes. | How are CMEs accelerated? |
| HXR imaging spectrometer | Sensitivity: >100 × RHESSI<br>Dynamic Range: >100:1 | Image electrons in the corona as they are accelerated. | How are electrons accelerated? |
| Gamma-ray imaging spectrometer | Angular resolution: 5 arcsec<br>Sensitivity: >10 × RHESSI | Image and separate all ion-associated sources. | How are ions accelerated? |
| ENA imaging spectrometer | Angular resolution: 0.1 degrees<br>Sensitivity: 10-100× better than STEREO/LET | | |
| EUV imaging spectrometer | Temperature: $10^4$–$10^7$ K<br>4D* imaging at <10 s cadence | Observe the coronal plasma where energy is released. | How is the plasma heated? |
| SXR imaging spectrometer | Energy res.: 1 eV (1 to 20 keV)<br>4D* imaging at <10 s cadence | | |

* "4D imaging" means imaging in two spatial dimensions over time and energy or wavelength.

# References

Abbett, W. P., & Hawley, S. L. 1999, Astrophysical Journal, 521, 906

Allred, J. C., Hawley, S. L., Abbett, W. P., and Carlsson, M. 2005, Astrophysical Journal, 630, 573

Caspi, A. and Lin, R. P. 2010, Astrophysical Journal Letters, in press.

Cohen, E., Kessler, D. A., & Levine, H. 2005, Physical Review E, 72, 66126

Druckmüller, M., Rušin, V., & Minarovjech, M. 2006, Contributions of the Astronomical Observatory Skalnaté Pleso, 36, 131

Emslie, A. G., Kucharek, H., Dennis, B. R., Gopalswamy, N., Holman, G. D., Share, G. H., Vourlidas, A., Forbes, T. G., Gallagher, P. T., Mason, G. M., Metcalf, T. R., Mewaldt, R. A., Murphy, R. J., Schwartz, R. A., & Zurbuchen, T. H. 2004, Journal of Geophysical Research, 109, 10104

Emslie, A. G., Dennis, B. R., Holman, G. D., and Hudson, H. S., 2005, Journal of Geophysical Research, 110, 11103

Forbes, T. G., et al. 2006, Space Science Reviews, 123, 251

Holman, G. D. et al., 2010, Space Science Reviews, in press.

Hurford, G. J., Krucker, S., Lin, R. P., Schwartz, R. A., Share, G. H., & Smith, D. M. 2006, Astrophysical Journal, 644, L93

Hurford, G. J., Schwartz, R. A., Krucker, S., Lin, R. P., Smith, D. M., & Vilmer, N. 2003, Astrophysical Journal, 595, L77

Kahler, S. W., & Ragot, B. R. 2008, Astrophysical Journal, 675, 846

Kontar, E. P. et al., 2010, Space Science Reviews, in press.

Krucker, S., Hurford, G. J., MacKinnon, A. L., Shih, A. Y., & Lin, R. P. 2008, Astrophysical Journal, 678, L63

Krucker, S., Hudson, H. S., Glesener, L., White, S. M., Masuda, S., Wuesler, J.-P., and Lin, R. P. 2010, Astrophysical Journal, 714, 1108.

Landi, E., Raymond, J. C., Miralles, M. P., & Hara, H. 2010, Astrophysical Journal, 711, 75

MacKinnon, A. L. 1989, Astronomy and Astrophysics (ISSN 0004-6361), 226, 284

Mewaldt, R. A., Leske, R. A., Shih, A. Y., Stone, E. C., Barghouty, A. F., Cohen, C. M. S., Cummings, A. C., Labrador, A. W., von Rosenvinge, T. T., & Wiedenbeck, M. E. 2010, Twelfth International Solar Wind Conference. AIP Conference Proceedings, 1216, 592

Mewaldt, R. A., Leske, R. A., Stone, E. C., Barghouty, A. F., Labrador, A. W., Cohen, C. M. S., Cummings, A. C., Davis, A. J., von Rosenvinge, T. T., & Wiedenbeck, M. E. 2009, Astrophysical Journal Letters, 693, L11

Qui, J. et al. 2004, Astrophysical Journal, 604, 900

Ramaty, R., and N. Mandzhavidze (2000), Gamma-rays from solar flares, in Highly Energetic Physical Processes and Mechanisms for Emission from Astrophysical Plasmas, ASP Conf. Ser., IAU Symp. 195, edited by P. C. H. Martens, S. Tsuruta, and M. A. Weber, p. 123

Share, G. H., Murphy, R. J., and Newton, E. K. 2001, Solar Physics, 201, 191

Shih, A. Y., Lin, R. P., & Smith, D. M. 2009, Astrophysical Journal Letters, 698, L152

Sui, L., & Holman, G. D. 2003, Astrophysical Journal, 596, L251

Tylka, A. J., & Lee, M. A. 2006, Astrophysical Journal, 646, 1319

Wang, T., Sui, L., and Qiu, J. 2007, Astrophysical Journal Letters, 661, L207

Yokoyama, T, Akita, K., Morimoto, T., Inoue, K., & Newmark, J., 2001, Astrophysical Journal Letters, 546, L69.

**Complementary White Papers**

Bandler et al. "XMS"

Christe et al. "The Focusing Optics Solar X-ray Imager"

Davila et al. "Understanding Magnetic Storage, Reconnection, and CME Initiation"

Doschek et al. "A Concept White Paper for a New Solar Flare Instrument Designed to Determine the Plasma Parameters in the Reconnection Region of Solar Flares at Flare Onset"

Doschek et al. "The High Resolution Solar-C International Collaboration"

Gary et al. "The Frequency Agile Solar Radiotelescope"

Gopalswamy et al. "Earth-Affecting Solar Causes Observatory (EASCO): A New View from Sun-Earth L5"

Laming et al. "Science Objectives for an X-Ray Microcalorimeter Observing the Sun"

McConnell et al. "Hard X-ray and gamma-ray polarization measurements of solar flares during the next solar maximum"

Shih et al. "Solar Ion Acceleration and the Flaring Atmosphere"

St. Cyr et al. "Solar Orbiter: Exploring the Sun-Heliosphere Connection"

Vourlidas et al. "Mission to the Sun-Earth L5 Lagrangian Point: An Optimal Platform for Heliophysics & Space Weather Research"